\documentclass[oribibl]{llncs}
\usepackage{amssymb}
\usepackage{epsfig,amsmath}
\usepackage{multicol}
\usepackage{latexsym}
\usepackage{smallsec}

\newcommand{\boxtheorem}{\hfill $\Box$}
\newcommand{\nit}[1]{{\it #1}}
\newcommand{\n}{~{\it not}~}
\newcommand{\sm}{\smallsetminus}

\newcommand{\tr}{\mathbf{t}}
\newcommand{\ta}{\mathbf{t_{a}}}
\newcommand{\fa}{\mathbf{f_{a}}}
\newcommand{\trs}{\tr^\star}

\newcommand{\tss}{\tr^{\star\star}}

\newcommand{\oa}{(\bar{a})}

\newcommand{\schn}{\models_{_N}}

\newcommand{\D}{\nit{D}}
\newcommand{\IC}{\nit{IC}}
\newcommand{\nn}{\nit{null}}

\newcommand{\oatstarr}{\!\!\_(\bar{a},\mathbf{t}^{\star\star})}

\newcommand{\dbic}{\D,\IC}

\newcommand{\mm}{\mathcal{M}}

\title{\vspace*{-1cm}{\bf Semantically Correct Query Answers in the Presence of Null Values}}

\author{{\bf Loreto Bravo}~~ \and~
 {\bf Leopoldo Bertossi}\\
Carleton
University\\ School of Computer Science\\Ottawa, Canada.\\
\{lbravo,bertossi\}@scs.carleton.ca}

\institute{}

\begin{document}
\pagestyle{plain} \thispagestyle{empty} \maketitle

\vspace{-8mm} \begin{abstract}\noindent For several reasons a
database may not satisfy a given set of integrity constraints
(ICs), but most likely most of the information in it is still
consistent with those ICs; and could be retrieved when queries are
answered. Consistent answers to queries wrt a set of ICs have been
characterized as answers that can be obtained from every possible
 minimally  repaired consistent version of the original
database. In this paper we consider databases that contain null
values and are also repaired, if necessary, using null values. For
this purpose, we propose first a precise semantics for IC
satisfaction in a database with null values that is compatible
with  the way null values are treated in commercial database
management systems. Next, a precise notion of repair is introduced
that privileges the introduction of null values when repairing
foreign key constraints, in such a way that these new values do
not create an infinite cycle of new inconsistencies. Finally, we
analyze how to specify this kind of  repairs of a database that
contains null values using disjunctive logic programs with stable
model semantics.
\end{abstract}

\vspace*{-3mm}
\section{Introduction}

In databases, integrity constraints (ICs) capture  the semantics
of the application domain, and help maintain the correspondence
between this  domain and  the database when updates are performed.
However, there are several reasons for a database to be or become
inconsistent wrt a given set of ICs \cite{book03}; and sometimes
 it could be difficult, impossible or undesirable
to repair the database in order to restore consistency
\cite{book03}. This process might be too expensive; useful data
might be lost;  it may not be clear how to restore the
consistency, and sometimes even impossible, e.g. in virtual  data
integration, where the access to the autonomous data sources may
be restricted \cite{bb-2005}.

In those situations, possibly most of the data is still consistent
and can be retrieved when queries are posed to the database. In
\cite{ABC99}, consistent data is characterized as the data that is
invariant under certain minimal forms of restoration of
consistency, i.e. as the data that is present in all minimally
repaired and consistent versions of the original instance, the
so-called  {\em repairs}. In particular, an answer to a query is
defined as consistent when it can be obtained as a standard answer
to the query from {\em every possible} repair.

More precisely, a repair of a database instance $D$, as introduced
in \cite{ABC99}, is a new instance of the same schema as $D$ that
satisfies the given ICs, and makes minimal under set inclusion the
symmetric set difference with the original instance, taken both
instances as sets of ground database atoms.

In \cite{ABC99,celle,cikm04,fuxSigmod05} algorithms and
implementations for consistent query answering (CQA) have been
presented, i.e. for retrieving consistent answers from
inconsistent databases. All of them work only with the original,
inconsistent database, without restoring its consistency. That is,
inconsistencies are solved at query time. This is in
correspondence with the idea that the above mentioned repairs
provide an auxiliary concept for defining the right semantics for
consistent query answers. However, those algorithms apply to
restricted classes of queries and constraints, basically those for
which the intrinsic complexity of CQA is still manageable
\cite{cm-2005}.

In \cite{tplp,greco03,padl,book03} a different approach is taken:
database repairs are specified as the stable models of disjunctive
logic programs, and in consequence consistent query answering
amounts to doing {\em cautious} or {\em certain} reasoning from
logic programs under the stable model semantics. In this way, it
is possible to handle any set of universal ICs and any first-order
query, and even beyond that, e.g. queries expressed in extensions
of Datalog. It is important to realize that the data complexity of
query evaluation in disjunctive logic programs with stable model
semantics \cite{voronkov} matches the intrinsic data complexity of
CQA \cite{cm-2005}, namely both of them are $\Pi^P_2$-complete.

All the previous work cited before did not consider the possible
presence of null values in the database, and even less their
peculiar semantics. Using null values to repair ICs was only
slightly considered in \cite{tplp,padl,book03}. This strategy to
deal with referential ICs seemed to be the right way to proceed
given the results presented in \cite{cali} that show that
repairing cyclic sets of referential ICs by introducing arbitrary
values from the underlying database domain leads to the
undecidability of CQA.

In \cite{cascon} the methodology presented in \cite{padl,book03},
based on specifying repairs using logic programs withe extra
annotation constants, was systematically extended in order to
handle both; (a) databases containing null values, and (b)
referential integrity constraints (RICs) whose satisfaction is
restored via introduction of null values. According to the notion
of IC satisfaction implicit in \cite{cascon}, those introduced
null values do not generate any new inconsistencies.

Here, we extend the approach and results in \cite{cascon} in
several ways. First, we give a precise semantics for integrity
constraint satisfaction in the presence of null values that is
both sensitive to {\em the relevance of the occurrence of a null
value} in a relation, and also compatible with the way null values
are usually treated in commercial database management systems (the
one given in \cite{cascon} was much more restrictive). The
introduced null values do not generate infinite repair cycles
through the same or other ICs, which requires a semantics for
integrity constraints satisfaction under null values that
sanctions that tuples with null values in attributes relevant to
check the IC  do not generate any new inconsistencies. A new
notion of repair is given accordingly. With the new repair
semantics CQA becomes decidable for a quite general class of ICs
that includes universal constraints, referential ICs, {\em NOT
NULL}-constraints, and foreign key constraints, even the cyclic
cases.

The logic programs that specify the repairs are modified wrt those
given introduced in \cite{cascon}, in such a way that the expected
one-to-one correspondence between the stable models and repairs is
recovered for {\em acyclic} sets of RICs.  Finally, we study
classes of ICs for which the specification can be optimized and a
lower complexity for CQA can be obtained.

\vspace{-3mm}
\section{Preliminaries} \label{sec:prelim}

\vspace{-1mm}
 We concentrate on relational databases, and we
assume we have a fixed relational schema $\Sigma=({\cal U}, {\cal
R}, {\cal B})$, where ${\cal U}$ is the possibly infinite database
domain such that $\nit{null} \in {\cal U}$, ${\cal R}$ is a fixed
set of database predicates, each of them with a finite, ordered
set of attributes, and ${\cal B}$ is a fixed set of built-in
predicates, like comparison predicates. $R[i]$ denotes the
attribute in position $i$ of predicate $R \in {\cal R}$. The
schema determines a language ${\cal L}(\Sigma)$ of first-order
predicate logic. A database instance $D$ compatible with $\Sigma$
can be seen as a finite collection of ground atoms of the form
$R(c_{1},...,c_{n})$,\footnote{Also called {\em database tuples}.
Finite sequences of constants in $\cal U$ are simply called {\em
tuples}.} where $R$ is a predicate in $\cal{R}$ and
$c_{1},...,c_{n}$ are constants in $\cal U$. Built-in predicates
have a fixed  extension in every database instance, not subject to
 changes. We need to define ICs because their syntax is
fundamental for what follows.

An {\em integrity constraint} is a sentence $\psi \in {\cal
L}(\Sigma)$ of the form:
\vspace*{-.3cm}\begin{equation}\label{eq:formatGen}
  \forall \bar{x}(\bigwedge_{i = 1}^{m} P_i(\bar{x}_i)
  ~\longrightarrow~\exists \bar{z} (\bigvee_{j=1}^n
  Q_{j}(\bar{y}_j,\bar{z}_j) \vee \varphi)),
\end{equation}

\vspace{-3mm} \noindent where $P_i, Q_j \in {\cal R}$, ~$\bar{x}=
\bigcup_{i = 1}^{m} \bar{x}_i$,~ $\bar{z}= \bigcup_{j = 1}^{n}
\bar{z}_j$, ~$\bar{y}_j ~\subseteq \bar{x}$, ~$\bar{x} \cap
\bar{z} = \emptyset$, ~$\bar{z}_i \cap \bar{z}_j=\emptyset$ for $i
\not= j$, and $m \geq 1$. Formula $\varphi$ is a disjunction of
built-in atoms from $\cal B$, whose variables appear in the
antecedent of the implication. We will assume that there is a
propositional atom ${\bf false} \in {\cal B}$ that is always false
in a database. Domain constants other than $\nn$ may appear
instead of some of the variables in a constraint of the form
(\ref{eq:formatGen}). When writing ICs, we will usually leave the
prefix of universal quantifiers implicit. A wide class of ICs can
be accommodated in this general syntactic class by appropriate
renaming of variables if necessary.

 A \emph{universal integrity constraint} (UIC) has the
form (\ref{eq:formatGen}), but with  $\bar{z}=
 \emptyset$, i.e. without existentially quantified variables:
 \vspace{-2mm}
\begin{equation}\label{eq:format}
  \bar{\forall}\bar{x}(\bigwedge_{i = 1}^{m} P_i(\bar{x}_i) ~\longrightarrow~
\bigvee_{j=1}^n Q_{j}(\bar{y}_j) \vee
    \varphi).
\end{equation}

\vspace{-3mm}\noindent A \emph{referential integrity constraint}
(RIC) is of the form (\ref{eq:formatGen}), but with $m=n=1$ and
$\varphi =\emptyset$, i.e. of the form\footnote{To simplify  the
presentation, we are assuming the existential variables appear in
the last attributes of $Q$, but they  may appear anywhere else in
$Q$.}: ~(here $\bar{x}' \subseteq \bar{x}$ and $P, Q \in {\cal
R}$)

\vspace{-2mm}
\begin{equation}\label{eq:formatRIC}
\forall \bar{x}~(P(\bar{x}) \longrightarrow \exists
\bar{y}~Q(\bar{x}',\bar{y})).
  \end{equation}
 Class (\ref{eq:formatGen}) includes  most ICs
commonly found in database practice, e.g.  a {\em denial
constraint} can be expressed as
  $\bar{\forall}\bar{x}(\bigwedge_{i = 1}^{m} P_i(\bar{x}_i) \longrightarrow
 \nit{\bf false}).$
Functional dependencies can be expressed by several implications
of the form (\ref{eq:formatGen}), each of them with a single
equality in the consequent. Partial inclusion dependencies are
RICs, and full inclusion dependencies are universal constraints.
We can also specify (single row) {\em check constraints} that
allow to express conditions on each row in a table, so they can be
formulated with one predicate in the antecedent of
(\ref{eq:formatGen}) and only a formula $\varphi$ in the
consequent. For example, $\forall xy(P(x,y) \rightarrow y>0)$ is a
check constraint.

In the following we will assume that we have a fixed finite set
$\IC$ of ICs of the form (\ref{eq:formatGen}). Notice that sets of
constraints of this form are always a consistent in the classical
sense, because empty database always satisfy them.

\vspace{-1mm}
\begin{example} For ${\cal R}= \{P,R,S\}$ and ${\cal
B}=\{>,=, {\bf false}\}$, the following are ICs:~ (a) $\forall
xyzw ~(P(x,y) \wedge R(y,z,w)  ~\rightarrow~ S(x) \vee (z \not =2
\vee w \leq y))$~ (universal).~ (b)$\forall xy (P(x,$ $y)
~\rightarrow$ $\exists z ~R(x,y,z))$~ (referential).~ (c)$\forall
x (S(x) ~\rightarrow~ \exists yz ( R(x,y) \vee R(x,y,z)))$.
\boxtheorem
\end{example}

\vspace{-1mm}\noindent Notice that defining $\varphi$ in
(\ref{eq:formatGen}) as a disjunction of built-in atoms is not an
important restriction, because an IC that has  $\varphi$ as a more
complex formula can be transformed into a set of constraints of
the form (\ref{eq:formatGen}). For example, the formula $\forall
xy\,(P(x,y) \rightarrow (x > y \vee (x =3 \wedge y=8)))$ can be
transformed into: $\forall xy\, (P(x,y)  \rightarrow (x > y \vee x
=3) )$ and $\forall xy\, (P(x,y) \rightarrow (x> y \vee y  =8) )$.

The \textit{dependency graph} $\cal{G}(\IC)$ \cite{sccc05} for a
set of ICs $\IC$ of the form (\ref{eq:formatGen}) is defined as
follows: Each database predicate $P$ in ${\cal R}$  appearing in
$\IC$ is a vertex, and there is a directed edge $(P_i, P_{j})$
from $P_i$ to $P_{j}$ iff there exists a constraint $ic \in \IC$
such that $P_i$ appears in the antecedent of $ic$ and $P_{j}$
appears in the consequent of $ic$.

\begin{example} \label{ex:graph2} For the set $\IC$ containing the
UICs $\nit{ic}_1\!: S(x) \rightarrow Q(x)$ and $\nit{ic}_2\!:$
$Q(x) \rightarrow R(x)$, and the RIC $\nit{ic}_3\!: Q(x)
\rightarrow \exists yT(x,y)$, the following is the dependency graph $\cal{G}(\IC)$:\\
\centerline{\includegraphics[width=3.5cm]{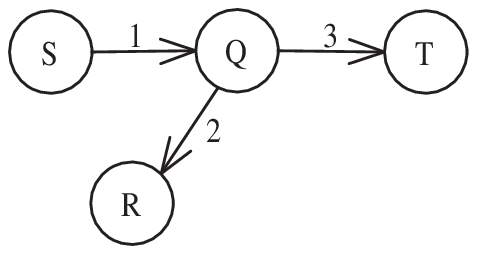}}

\noindent the edges are labelled just for reference. Edges $1$ and
$2$ correspond to the constraints $\nit{ic}_1$ and $\nit{ic}_2$,
resp., and edge $3$  to $\nit{ic}_3$. \boxtheorem
\end{example}

\vspace{-2mm}\noindent A {\em connected component} in a graph is a
maximal subgraph such that for every pair $(A$, $B)$ of its
vertices, there is a path from $A$ to $B$ or from $B$ to $A$. For
a graph $\cal G$, ${\cal C(G)}:=\{c ~|~ c$ ~is a connected
component in ${\cal G}\}$; and ${\cal V}(\cal G)$ is the set of
vertices of $\cal G$.

\begin{definition}\label{de:RICacyc}\em Given a set $\IC$ of
UICs and RICs, $\IC_{\!U}$ denotes the set of UICs in $\IC$. ~ The
{\em contracted dependency graph}, ${\cal G}^C(\IC)$, of $\IC$ is
obtained from ${\cal G}(\IC)$ by replacing, for every $c \in {\cal
C}({\cal G}(\IC_{\!U}))$,\footnote{Notice that for every $c \in
{\cal C}({\cal G}(\IC_{\!U}))$, it holds  $c \in {\cal C}({\cal
G}(\IC))$.} the vertices in ${\cal V}(c)$ by a single vertex and
deleting all the edges associated to the elements of $\IC_{\!U}$.
Finally, $\IC$  is said to be {\em RIC-acyclic} if ${\cal
G}^C(\IC)$ has no cycles. \boxtheorem
\end{definition}

\begin{example} \label{ex:graph3}(example \ref{ex:graph2} cont.)
The contracted dependency graph ${\cal G}^C(\IC)$ is obtained by
replacing in ${\cal G}(\IC)$ the edges $1$ and $2$ and their end
vertices by a vertex labelled with $\{Q,R,S\}$. 

\centerline{\includegraphics[width=2.7cm]{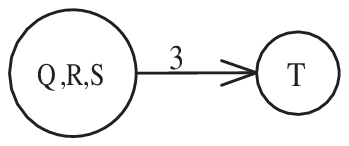}} \noindent
Since there are no loops in ${\cal G}^C(\IC)$,  $\IC$ is
RIC-acylic.
\noindent If we add a new UIC: $T(x,y) \rightarrow R(y)$ to $\IC$,
all the vertices belong to the same connected component. ${\cal
G}(\IC)$ and ${\cal G}^C(\IC)$ are, respectively:
\vspace*{-3mm}\begin{multicols}{2}
\centerline{\includegraphics[height=1.95cm]{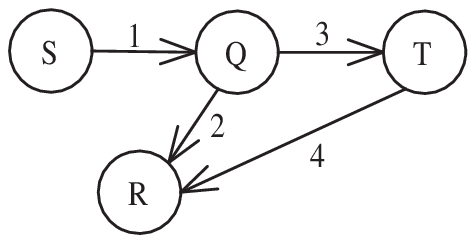}}\centerline{\includegraphics[height=1.9cm]{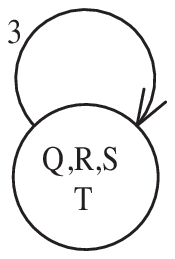}}
\end{multicols}\vspace*{-3mm}\noindent Since there is a self-loop in ${\cal G}^C(\IC)$, the new
$\IC$
is {\em not} RIC-acylic. \boxtheorem
\end{example}
\vspace{-2mm}As expected, a set of UICs is always RIC-acyclic.

\section{IC Satisfaction in Databases with Null Values}\label{sec:ICSat}

We deal with incomplete databases in the classic sense that some
information is represented using null values \cite{IL84} (cf. also
\cite{grahne91}). More recently, the notion of incomplete database
has been used in the context of virtual data integration
\cite{lenzerini,bb-2005}, referring to data sources that contain a
subset of the data of its kind in the global system; and in
inconsistent databases \cite{cali,cm-2005}, referring  to the fact
that inconsistencies may have occurred due to missing information
and then, repairs are obtained through insertion of new tuples.

There is no agreement in the literature on the semantics of null
values in relational databases. There are several different
proposals in the research literature
\cite{reiter84,AM86,LL94,Lie82}{}, in the SQL standard
\cite{TG01,SQL2003},  but also implicit semantics in the different
ways null values are handled in commercial database management
systems (DBMSs).

Not even within the SQL standard there is a homogenous and global
semantics of integrity constraint satisfaction in databases with
null values; rather, different definitions of satisfaction are
given for each type of constraint. Actually, in the case of
foreign key constraints, three different semantics are suggested
({\em simple-}, {\em partial-} and {\em full-match}). Commercial
DBMSs implement only the  simple-match semantics for foreign key
constraints.

One of the reasons why it is difficult to agree on a semantics is
that a null value can be interpreted as an unknown, inapplicable
or even withheld value. Different null constants can be used for
each of these different interpretations \cite{libkin95}. Also the
use of more than one null value (of the same kind), i.e. labelled
nulls, has been suggested \cite{reiter86}, but in this case every
new null value uses a new fresh constant; for which   the {\em
unique names assumption} does not apply. The latter alternative
allows to keep a relationship between null values in different
attributes or relations. However commercial DBMSs consider only
one null value, represented by a single constant, that can be
given any of the interpretations mentioned above.

 In \cite{cascon} a semantics for null values was
adopted, according to which a tuple with a null value in any of
its attributes would not be the cause for any inconsistencies. In
other words, it would not be necessary to check tuples with null
values wrt possible violations  of ICs (except for {\em NOT NULL}-
constraints, of course). This assumption is consistent in some
cases with  the practice of DBMSs, e.g. in IBM DB2 UDB. Here we
will propose a semantics that is less liberal in relation to the
participation of null values in inconsistencies; a sort of
compromise solution considering the different alternatives
available.

\vspace{-2mm}\begin{example} For $\IC$ containing only
$\nit{\psi}_1\!: P(x,y,z) \rightarrow R(y,z)$, the database
$\D=\{P(a,b,null)\}$ would be: (a) Consistent wrt the semantics in
\cite{cascon} because there is a null value in the tuple (b)
Consistent wrt the simple-match semantics of SQL:2003
\cite{SQL2003}, because there is a null value in one of the
attributes in the set $\{P[2],P[3],R[1],R[2]\}$ of attributes that
are relevant to check the constraint. (c) Inconsistent wrt the
partial-match semantics in SQL:2003, because there is no tuple in
$R$ with a value $b$ in its first attribute. (d) Inconsistent wrt
the full-match semantics in SQL:2003, because there cannot be a
$\nn$ in an attribute that is referencing a different table.

If we consider, instead of $\nit{\psi}_1$, the constraint
$\nit{\psi}_2\!: P(x,y,z) \rightarrow R(x,y)$, the same database
would be consistent only for the semantics in  \cite{cascon},
because the other semantics consider only the null value in the
attributes that are relevant to check the constraint, and in this
case there is no null value there. \boxtheorem
\end{example}
We want a null-value semantics that generalizes the semantics
defined in SQL:2003 \cite{SQL2003} and is used by DBMSs, like IBM
DB2 UDB. For this reason we consider only one kind of null value,
that is interpreted in the same way for different types of ICs. We
also want our null-value semantics to be uniform for a wide class
of ICs, not only for the type of constraints supported commercial
DBMS.

\begin{example}\label{ex:FK} Consider a database with a table that stores courses with the
professor that taught it and the term, and a table that stores the
experience of each professor in each course with the number of
times (s)he has taught the course. We have a foreign key
constraint based on the  RIC $\forall xyz (Course(x,y,z)$
$\rightarrow$ $ \exists w $ $Exp(y,$ $x,$ $w))$ together with the
constraint expressing that table {\it Exp} has $\{ \it ID, Code\}$
as a key. We can be sure there are no null values in those two
attributes. Now consider the instance $D$:
\begin{center}
\begin{tabular}{c|c|c|c|}
\hline ~~Course ~& ~Code~ & ~ ID~ & ~Term~\\ \hline
 & {\em CS27} & {\em 21} & W04  \\
 & {\em CS18} &{\em 34} & {\em \nn}\\
 & {\em CS50} &{\em {\nn}}& W05   \\
 \cline{2-4}
 \end{tabular}
~~~\begin{tabular}{c|c|c|c|}
\hline ~~Exp ~& ~\underline{ID}~ & ~\underline{Code}~ & ~Times~\\
\hline
 & {\em 21} & {\em CS27} & 3\\
 & {\em 34} &  {\em CS18}  & {\em {\nn}}  \\
 & {\em 45} &  {\em CS32} & {\em 2} \\
 \cline{2-4}
 \end{tabular}
 \end{center}
In IBM DB2, this database is accepted as consistent. The null
values in columns {\it Term} and {\it Times} are not relevant to
check the satisfaction of the constraints. In order to check the
constraint the only attributes that we need to pay attention to
are {\it ID} and {\it Code}. If $\nn$ is in the one of these
attributes in table {\it Course}, the tuple is considered to be
consistent, without checking table {\it Exp}. For example {\it
Course(CS50,\nn,W05)} has a null value in {\it ID}, therefore DB2
does not check if there is a tuple in {\it Exp} that satisfies the
constraint. It does not even check that there exists a tuple in
{\it Exp} with attribute {\it Code=CS50}.

This behavior for foreign key constraints  is called simple-match
in the SQL standard, and is the one implemented in all commercial
DBMS. The partial- and full-match would not accept the database as
consistent, because partial-match would require {\it Exp} to have
a tuple ~$(\mbox{any non-null value},34, \mbox{any value})$; and
full-match would not allow a tuple with $\nn$ in attributes {\it
ID} or {\it Code} in table {\it Course}.

If we try to insert tuple  {\it(CS41,18, {\nn})} into table {\it
Course}, it would be rejected by DB2. This is because the
attributes {\it ID} and {\it Code} are relevant to check the
constraint and are different from \nn, but there is no tuple in
 {\it Exp} with {\it ID=18} and {\it Code=CS41}. \boxtheorem
\end{example}

\begin{example}\label{ex:check} Consider the single-row check constraint
 $\forall {\it ID}\,\forall{\it Name}\,\forall{\it
Salary}\, (\nit{Emp}$ $({\it ID},$ ${\it Name},$ ${\it Salary})$
$\rightarrow $ $ {\it Salary} >100) $ and the database $D$ below.
DB2 accepts
\vspace{-3mm} \begin{multicols}{2}
\begin{tabular}{c|c|c|c|}
\hline ~~Emp~ & ~ID~ & ~Name~ & ~Salary~\\ \hline
 & {\em 32} & {\em \nn}&{\em 1000}\\
  & {\em 41} & {\em Paul}&{\em \nn}\\
 \cline{2-4}
 \end{tabular}

\noindent this database instance as consistent. Here, in order to
check the satisfaction of the constraint, we only need to verify
\end{multicols} \vspace{-4mm} \noindent  that the attribute {\it Salary}
is bigger than 100; therefore the only attribute that is relevant
to check the constraint is {\it Salary}. DBMSs will accept as
consistent any state where the condition (the consequent)
evaluates to {\it true} or {\it unknown}. The latter is the case
here. Tuple {\emph (32, \nn, 50)} could not be inserted because
$Salary >100$ evaluates to {\it false}. Notice that the null
values in attributes other that {\it Salary} are not even
considered in the verification of the satisfaction. \boxtheorem
\end{example}
When dealing with primary keys, DBMSs use a bag semantics instead
of the set semantics, that is, a table can have two copies of the
same tuple. The following example illustrates the issue.

\begin{example}\label{ex:PK} Since the SQL standard allows duplicate rows,
i.e.  uses the bag semantics, it is possible to have the database
$D$ below. If this database had $P[1]$
\vspace{-5mm}\begin{multicols}{2}\begin{center}
  \begin{tabular}{c|c|c|} \hline
    ~~~P~~&~A~&~B~\\
    \hline  & a & b\\
      & a & b\\
      \cline{2-3}
  \end{tabular}
\end{center}
\vspace*{-6mm}as the primary key, then $D$ would not have been
accepted as a consistent state, i.e. the insertion of the second
tuple $P(a,b)$ would have been rejected.
\end{multicols} \vspace{-3mm} \noindent This is one of the cases in which the SQL standard
deviates from the relational model, where duplicates of a row are
not considered. In a commercial DBMS a primary key is checked  by
adding an index to the primary key and then ensuring that there
are no duplicates. Therefore if we try to check the primary key by
using the associated functional dependency $P(x,y), P(x,z)
\rightarrow y = z$ we would not have the same semantics since $D$
satisfies the functional dependency  in this classical,
first-order representation. \boxtheorem
\end{example}
With the type of first-order constraints that we are considering,
we cannot enforce a bag semantics, therefore we will assume that
$D$ is consistent.

In order to develop a null-value semantics that goes beyond the
ICs supported by DBMSs,  we analyze other examples.

\begin{example}\label{ex:check-multi-row}Consider the UIC $\forall xyzstuw(Person(x,y,z,w)$ $\wedge$ $ Person
(z,s,t,u)$ $\rightarrow ~~$ $u> w+15)$,\, and the database $D$
below. This constraint can be considered
\vspace{-7mm}\begin{multicols}{2}\begin{center}
\begin{tabular}{c|c|c|c|c|}
\hline ~~Person~ & ~Name~ & ~Dad~ & ~Mom~ & ~Age~\\ \hline
 & {\em Lee} & {\em Rod} &{\em  Mary}& {\em 27} \\
  & {\em Rod} & {\em Joe} & {\em Tess} & {\em 55} \\
  & {\em Mary} & {\em Adam }& {\em Ann} & {\em \nn}\\
 \cline{2-5}
 \end{tabular}\end{center}
\vspace*{-3.5mm} as a multi-row check constraint. If we want to
naturally extend the semantics for single-row check constraints,\,
$D$ would\, be consistent\, iff\, the\, condition
\end{multicols} \vspace*{-3mm}\noindent
evaluates to {\it true} or {\it unknown}. In this case, $D$ would
be consistent because the condition evaluates to {\it unknown} for
$u=\nn$ and $w=27$. Here the relevant attributes to check the IC
are {\it Name, Mom, Age}. \boxtheorem
\end{example}

\begin{example} \label{ex:IND} Consider the UIC $\forall xyz (Course(x,y,z)\rightarrow
Employee(y,z))$ and the database $D$:
\vspace{-5mm}\begin{center}~~~~~~
\begin{tabular}{c|c|c|c|}
\hline ~~Course ~& ~Code~ & ~Term~ & ~ID~\\ \hline
 & {\em CS18} & {\em W04}&{\em 34}\\
 \cline{2-4}
 \end{tabular}
~~~\begin{tabular}{c|c|c|} \hline ~~Employee~ &~ Term~ & ~ID~\\
\hline
 & {\em W04} & {\em \nn}\\
 \cline{2-3}
 \end{tabular}
 \end{center}
\vspace{-2mm}Since {\it Term, ID} is not a primary key of  {\it
Employee}, the constraint is not a foreign key constraint, and
therefore it is not supported by commercial DBMS. In contrast to
foreign key constraints, now we can have a null value in the
referenced attributes.

In order to extend the semantics used in commercial DBMS. to this
case, we refer to the literature. For example, in \cite{LL94} the
satisfaction of this type of constraints is defined as follows: An
IC ~$\forall \bar{x}\bar{y}{P(\bar{x})\rightarrow \exists \bar{z}
Q(\bar{y}, \bar{z})}$~ is satisfied if, for every tuple ${t_1 \in
P}$, there exists a tuple ${t_2 \in Q}$, such that $t_1$ provides
{\it less or equal information} than $t_2$,  i.e.  for every
attribute, the value in $t_1$ is the same as in $t_2$ or the value
in $t_1$ is $\nn$.

In this example we have the opposite situation: ({\em W04},{\em
34}) does {\em not} provide less or equal information than ({\em
W04},{\em \nn}). Therefore, we consider the database to be
 {inconsistent} wrt the constraint. Note that the only attributes
 that are relevant to check the constraint are {\it Term} and
 {\it ID}. \boxtheorem
\end{example}
Examples \ref{ex:check}, \ref{ex:FK}, \ref{ex:check-multi-row} and
\ref{ex:IND} show that there are some attributes that are
``relevant" when the satisfaction of a constraint is checked
against a database.
\begin{definition} \em \label{def:rel} For $t$ a term, i.e. a variable or a domain constant, let   $\nit{pos}^R\!(\psi,t)$ be the set of positions in
predicate $R \in {\cal R}$ where $t$ appears in $\psi$. The
set ${\cal A}$ of {\em relevant attributes} for an IC $\psi$ of the form (\ref{eq:formatGen}) is\\
${\cal A}(\psi)=\{R[i] ~|~ x$ is variable present at least twice
in $\psi, \mbox{ and } i \in  \nit{pos}^R\!(\psi,x)\} ~\cup $\\
\hspace*{1.3cm}$\{R[i] ~|~ c$ is a constant in $\psi$ and $i \in
\nit{pos}^R\!(\psi,c)\}$. \boxtheorem
\end{definition}
Remember that $R[i]$ denotes a position (or the correspondent
attribute) in relation $R$. In short, the relevant attributes for
a constraint are those involved in joins, those appearing both in
the antecedent and consequent of (\ref{eq:formatGen}), and those
in $\varphi$.

\begin{definition} \em For a set of attributes ${\cal
A}$ and a predicate $P \in {\cal R}$, we denote by $P^{\cal A}$
the predicate $P$ restricted to the attributes in ${\cal A}$.
$\D^{\cal A}$ denotes the database $\D$ with all its database
atoms projected onto the attributes in ${\cal A}$, i.e. $D^{\cal
A} = \{P^{\cal A}(\Pi_{\cal A}(\bar{t})) ~|~ P(\bar{t}) \in D\}$,
where $\Pi_{\cal A}(\bar{t})$ is the projection on ${\cal A}$ of
tuple $\bar{t}$. ~$D^{\cal A}$ has the same underlying domain
${\cal U}$ as $D$. \boxtheorem
\end{definition}

\vspace*{-3mm}\begin{example} Consider a UIC $\psi:~ \forall
xyz(P(x,y,z) \rightarrow R(x,y))$ and $D$ below.
\vspace{-3mm}\begin{multicols}{2}\begin{center}\begin{tabular}{c|c|c|c|}
  \hline
  ~~~P~ & ~A~ & ~B~ & ~C~ \\
  \hline
   & a & b & a \\
     & b & c & a \\
       \cline{2-4}
\end{tabular}
~~~
\begin{tabular}{c|c|c|}
  \hline
  ~~~R~ & ~A~ & ~B~  \\
  \hline
   & a & 5  \\
     & a & 2  \\
       \cline{2-3}
\end{tabular}\end{center}\vspace{-4mm}Since $x$ and $y$ appear twice in $\psi$, ${\cal
A}(\psi)$ $=$ $\{P[1],R[1],P[2],R[2]\}$. The value in $z$\,
should\, not\, be relevant\, to check the \end{multicols}
\vspace{-4mm} \noindent
 satisfaction of the constraint, because we only want to make
sure that the values in the first two attributes in $P$ also
appear in $R$. Then, checking this is equivalent to checking if
$\forall xy(P^{{\cal A}(\psi)}(x,y) \rightarrow R^{{\cal
A}(\psi)}(x,y))$ is satisfied by $\D^{{\cal A}(\psi)}$. For a more
complex constraint, such as $\gamma\!:~ \forall xyzw(P(x,y,z)
\wedge R(z,w) ~\rightarrow$ $\exists v R(x,v) \vee w>3)$, variable
$x$ is relevant to check the implication, $z$ is needed to do the
join, and $w$ is needed to check the comparison, therefore ${\cal
A}(\gamma)= \{P[1],R[1],P[3],R[2]\}$.\\

\vspace{-3mm}\noindent $\D^{{\cal A}(\psi)}:$ \hspace{4.7cm}
$\D^{{\cal A}(\gamma)}:$

\noindent \begin{tabular}{c|c|c|}
  \hline
  ~~P$^{{\cal A}(\psi)}$~ & ~A~ & ~B~ \\
  \hline
   & a & b  \\
     & b & c \\
       \cline{2-3}
\end{tabular}
~~~
\begin{tabular}{c|c|c|}
  \hline
  ~~R$^{{\cal A}(\psi)}$~ & ~A~ & ~B~  \\
  \hline
   & a & 5  \\
     & a & 2  \\
       \cline{2-3}
\end{tabular}~~~
\begin{tabular}{c|c|c|}
  \hline
  ~~P$^{{\cal A}(\gamma)}$~ & ~A~ & ~C~ \\
  \hline
   & a & a  \\
     & b & a \\
       \cline{2-3}
\end{tabular}
~~~
\begin{tabular}{c|c|c|}
  \hline
  ~~R$^{{\cal A}(\gamma)}$~ & ~A~ & ~B~  \\
  \hline
   & a & 5  \\
     & a & 2  \\
       \cline{2-3}
\end{tabular}

\boxtheorem
\end{example}
An important observation we can make from Examples \ref{ex:check},
\ref{ex:FK}, \ref{ex:check-multi-row} and \ref{ex:IND} is that,
roughly speaking, a constraint is satisfied if any of the relevant
attributes  has a $\nn$  or the constraint is satisfied in the
traditional way (i.e. first-order satisfaction and null values
treated as any other constant). We introduce a special predicate
$\nit{IsNull}(\cdot)$, with $\nit{IsNull}(c)$ true iff $c$ is
$\nn$, instead of using the built-in comparison atom $c=\nn$,
because in traditional DBMS this equality would be always
evaluated as $\nit{unknown}$ (as observed in \cite{reiter84}, the
{\em unique names assumption} does not apply to null values).

 \vspace*{-2mm}\begin{definition} \em \label{def:pseudosat}
A constraint $\psi$ as in (\ref{eq:formatGen}) is satisfied in the
database instance $D$, denoted $D \models_{_N} \psi$ iff $D^{{\cal
A}(\psi)} \models \psi^N$, where $\psi^N$ is \vspace{-4mm}
\begin{equation}\label{eq:Satrule2}
\forall \bar{x}(\bigwedge_{i = 1}^{m} P_i^{{\cal
A}(\psi)}(\bar{x}_i)
   ~\rightarrow~ (\!\!\!\bigvee_{v_j \in {\cal A}(\psi) \cap \bar{x}} \!\!\!\!{\it
IsNull}(v_j) ~\vee~ \exists  \bar{z} (\bigvee_{j=1}^n
   Q_{j}^{{\cal A}(\psi)}(\bar{y}_j,\bar{z}_j) ~\vee~
    \varphi)
  )),
\end{equation}

\vspace{-3mm}\noindent where $\bar{x}= \cup_{i=1}^m \bar{x}_i$ and
$\bar{z}= \cup_{j=1}^n \bar{z}_j$.  $D^{{\cal A}(\psi)} \models
\psi^N$ refers to classical first-order satisfaction where
 $\nn$ is treated as any other constant in $\cal U$. \boxtheorem
\end{definition}
We can see from Definition \ref{def:pseudosat} that there are
basically two cases for constraint satisfaction: (a) If there is a
$\nn$ in any of the relevant attributes in the antecedent,  then
the constraint is satisfied. (b) If no null values appear in them,
then the second disjunct in  the consequent of formula
(\ref{eq:Satrule2}) has to be checked, i.e, the consequent of the
original IC restricted to the relevant attributes. This can be
done as usual, treating nulls as any other constant.

Formula (\ref{eq:Satrule2}) is a direct translation of formula
(\ref{eq:formatGen}) that keeps the relevant attributes. In
particular, if the original constraint is universal, so is the
transformed version. Notice that the transformed constraint is
domain independent, and then its satisfaction can be checked by
restriction to the active domain.

As mentioned before, the semantics for IC satisfaction introduced
in \cite{cascon} considered that tuples with $\nn$ never generated
any inconsistencies, even when the null value was not in a
relevant attribute. For example, under the semantics in
\cite{cascon}, the instance $\{P(b,\nn)\}$ would be consistent wrt
the IC $\forall xy (P(x,y) \rightarrow R(x))$, but it is
intuitively clear that there should be a tuple $R(b)$. The new
semantics corrects this, and adjusts to the semantics implemented
in commercial DBMS.

Notice that in a database without null values, Definition
\ref{def:pseudosat} (so as the definition in \cite{cascon})
coincides with the traditional, first-order definition of IC
satisfaction.

\begin{example} Given the ICs:
(a) $\forall xyz(P(x,y,z) \rightarrow R(x,y))$, (b) $\forall
x(T(x)\rightarrow \exists yz P(x,$ $y,z))$, the database instance
$D$ below  is consistent.
\begin{center}\begin{tabular}{c|c|c|c|}
  \hline
  ~~~P~ & ~A~ & B & ~C~ \\
  \hline
   & a & d & e \\
     & b &~\nn~  & g \\
       \cline{2-4}
\end{tabular}
~~~~~~~~
\begin{tabular}{c|c|c|}
  \hline
  ~~~R~ & ~D~ & ~E~  \\
  \hline
   & a & d  \\
     &  &   \\
       \cline{2-3}
\end{tabular}
~~~~~~~~
\begin{tabular}{c|c|}
  \hline
  ~~~T~ & ~F~ \\
  \hline
   & b  \\
     &    \\
       \cline{2-2}
\end{tabular}
\end{center}
For (a), the variables $x$ and $y$ are relevant to check the
constraint, therefore ${\cal A}_1 = \{P[1],$ $R[1],$ $P[2],$
$R[2]\}$; and for (b), the variable $x$ is relevant to check the
constraint; therefore ${\cal A}_2 = \{P[1],T[1]\}$.

\noindent $D^{{\cal A}_1}:$ \hspace{5.4cm} $D^{{\cal A}_2}:$

\vspace{-1mm}\noindent\begin{center} \begin{tabular}{c|c|c|}
  \hline
  ~~P$^{{\cal A}_1}$~ & ~A~ & B  \\
  \hline
   & a & d  \\
     & b & $\nn$ \\
       \cline{2-3}
\end{tabular}
~~~
\begin{tabular}{c|c|c|}
  \hline
  ~~R$^{{\cal A}_1}$~ & ~D~ & ~E~  \\
  \hline
   & a & d  \\
     &  &   \\
       \cline{2-3}
\end{tabular}~~~~~~~~~~~~\begin{tabular}{c|c|}
  \hline
  ~~P$^{{\cal A}_2}$~ & ~A~   \\
  \hline
   & a   \\
     & b   \\
       \cline{2-2}
\end{tabular}
~~~
\begin{tabular}{c|c|}
  \hline
  ~~T$^{{\cal A}_2}$ ~& ~F~ \\
  \hline
   & b  \\
     &    \\
       \cline{2-2}
\end{tabular}\end{center}
To check if $D \schn$ $\forall xyz(P(x,y,z)$ $ \rightarrow$ $
R(x,y))$, we need to check if $D^{{\cal A}_1} \models$ $\forall
xy(P^{{\cal A}_1}(x,y) \rightarrow (\nit{IsNull}(x) \vee
\nit{IsNull}(y) \vee R^{{\cal A}_1}(x,y)))$ For $x=a$ and $y=d$,
$D^{{\cal A}_1} \models P^{{\cal A}_1}(a,d)$, but none of them is
a null value, i.e. $\nit{IsNull}(a)$ and $\nit{IsNull}(d)$ are
both false, therefore we need to check if $D^{{\cal A}_1} \models
R^{{\cal A}_1}(a,d)$. For $x=b$ and $y=\nn$, $D^{{\cal A}_1}
\models P^{{\cal A}_1}(b,\nn)$, and since $D^{{\cal A}_1} \models
\nit{IsNull}(\nn)$, the constraint is satisfied. The same analysis
can be done to prove that $D$ satisfies constraint (b), this is by
checking $D^{{\cal A}_2} \models$ $\forall x(T^{{\cal A}_2}(x)
\rightarrow (\nit{IsNull}(x)$ $\vee P^{{\cal A}_2}(x)))$

If we add tuple $P(f,d,\nn)$ to $D$,  it would become inconsistent
wrt constraint (a), because $D^{{\cal A}_1} \not \models$
$(P^{{\cal A}_1}(f,d) \rightarrow (\nit{IsNull}(f) \vee
\nit{IsNull}(d) \vee R^{{\cal A}_1}(f,d)))$.\boxtheorem
\end{example}

\begin{example} Consider the IC $\psi$:~ $\forall x y w z~((P_1(x,y,w) \wedge
P_2(y,z)) \rightarrow \exists u~ Q(x,z,u)) $ and the database $D$:
\begin{center}
\begin{tabular}{c|c|c|c|}
\hline ~~~P$_1$~~ & ~A~ & ~B~ & ~C~\\ \hline
 & {\em a} & {\em b}&{\em c}\\
& {\em d} & ~{\em {\nn}}~ & {\em c}  \\
& {\em b} & {\em e} &~{\em {\nn}}~ \\
& ~{\em {\nn}}~ & {\em b} & {\em b}\\
 \cline{2-4}
 \end{tabular}
~~\begin{tabular}{c|c|c|} \hline ~~~P$_2$~~ & ~D~ & ~E~\\
\hline
 & {\em b} & {\em a}\\
  & {\em e} & {\em c}\\
 & {\em d} & ~{\em \nn}~\\
  & ~{\em \nn}~ & {\em b}\\
 \cline{2-3}
 \end{tabular}~~
\begin{tabular}{c|c|c|c|}
\hline ~~~$Q$~~ & ~F~ & ~G~ & ~~H~~\\ \hline
 & {\em a} & {\em a}&{\em c}\\
& {\em b} & ~{\em {\nn}}~ & {\em c}  \\
& {\em b} & {\em c} &{\em d} \\
& ~{\em {\nn}}~ & {\em c} & {\em a}\\
 \cline{2-4}
 \end{tabular} \end{center}
Variables $x$, $y$ and $z$ are relevant to check the constraint,
therefore the set of relevant attributes is ${\cal A}(\psi)=
\{P_1[1],P_1[2],P_2[1],P_2[2],Q[1],Q[2]\}$. Then we need to check
if $D^{{\cal A}(\psi)} \models \forall x y z~((P_1^{{\cal
A}(\psi)}(x,y)$ $\wedge$ $P_2^{{\cal A}(\psi)}(y,z))$
$\rightarrow$ $(\nit{IsNull}(x) \vee$ $\nit{IsNull}(y) \vee$
$\nit{IsNull}(z) \vee Q^{{\cal A}(\psi)}(x,z)) $, where $\D^{{\cal
A}(\psi)}$ is
\begin{center}
\begin{tabular}{c|c|c|}
\hline ~~$P_1^{{\cal A}(\psi)}~$ & A & B \\ \hline
 & {\em a} & {\em b}\\
& {\em d} & ~{\em {\nn}}~\\
& {\em b} & {\em e} \\
& ~{\em {\nn}}~ & {\em b} \\
 \cline{2-3}
 \end{tabular}
~~~~\begin{tabular}{c|c|c|} \hline ~~$P_2^{{\cal A}(\psi)}~$ & D & E\\
\hline
 & {\em b} & {\em a}\\
  & {\em e} & {\em c}\\
 & {\em d} & ~{\em \nn}~\\
  & ~{\em \nn}~ & {\em b}\\
 \cline{2-3}
 \end{tabular}~~~~
\begin{tabular}{c|c|c|}
\hline ~~$Q^{\!{\cal A}(\psi)}$~ & F & G\\ \hline
 & {\em a} & {\em a}\\
& {\em b} & ~{\em {\nn}}~ \\
& {\em b} & {\em c} \\
& ~{\em {\nn}}~ & {\em c} \\
 \cline{2-3}
 \end{tabular} \end{center}
When checking the satisfaction of $D^{{\cal A}(\psi)} \models
\psi^{N}$, $\nn$ is treated as any other constant. For example for
$x=d$, $y=\nn$ and $z=b$, the antecedent of the rule is satisfied
since $P_1^{{\cal A}(\psi)}(d,\nn)\in D^{\cal A}$ and $P_2^{{\cal
A}(\psi)}(\nn,a) \in D^{\cal A}$. If $\nn$ had been treated as a
special constant, with no unique names assumption applied to it,
the antecedent would have been false. For these values  the
consequence is also satisfied, because $\nit{IsNull}(\nn)$ is
true. In this example, $D^{{\cal A}(\psi)} \models \psi^{N}$, and
the database satisfies the constraint. \boxtheorem
\end{example}

 \noindent Notice that in order for formula (\ref{eq:Satrule2}) to have $\bar{z} \not =
  \emptyset$, i.e. existential quantifiers, there must exist an atom $Q_j(\bar{y}_j,\bar{z}_j)$ in
  the corresponding IC of the form (\ref{eq:formatGen}), such that $\bar{z}_j$ has a
  repeated variable. This is because that is the only case in
  which a constraint can have $({\cal A}(\psi) \sm \bar{x}) \not =
  \emptyset$.

\begin{example}
  Given $\psi\!: \forall x (P(x,y) \rightarrow \exists z Q(x,z,z))$ and
  $D=\{P(a,b),$ $P(\nn,c),$ $Q(a,\nn,\nn)\}$,
  ${\cal A}(\psi)=\{P[1],$ $Q[1],$ $Q[2],$ $Q[3]\}$.
  $D$ satisfies $\psi$ iff $D^{{\cal A}}\models \psi^N$, with $D^{{\cal
  A}(\psi)}=\{P^{\cal A}(a), P^{\cal A}(\nn),Q^{\cal A}(a,\nn,\nn)\}$
and  $\psi^N\!:
  \forall x (P^{{\cal A}(\psi)}(x) \rightarrow (\nit{IsNull}(x)$ $\vee$
  $\exists z Q^{{\cal A}(\psi)}(x,z,z)))$. The constraint is satisfied, because for $x=a$ it is
 satisfied given that there exists the satisfying value $\nn$ for $z$; and for
 $x=\nn$ the constraint is satisfied given that
  $\nit{IsNull}(\nn)$ is true. \boxtheorem
\end{example}

\noindent  The predicate ${\it IsNull}$ also allows us to specify
{\em NOT NULL}-constraints, which are common in commercial DBMS,
and prevent certain attributes from taking a null value. As
discussed before, this constraint is different from having $x \not
= null$.

\begin{definition}\label{def:NNC}\em A {\em NOT~NULL}-constraint (NNC) is a denial
constraint of the form \vspace{-2mm}
\begin{equation}\label{eq:SatruleNNC}
  \bar{\forall} \bar{x}(P(\bar{x}) \wedge {\it IsNull}(x_i)
  \rightarrow {\bf false}),
\end{equation}
where $x_i \in \bar{x}$ is in the position of the attribute that
cannot take null values. For a NNC $\psi$, we define $D \schn
\psi$ iff $D \models \psi$ in the classical sense, treating $\nn$
as any other constant. \boxtheorem \end{definition}

\noindent Notice that a NNC is not of the form
(\ref{eq:formatGen}), because it contains the constant $\nn$. This
is why we give a separate definitions for them. By adding NNCs we
are able to represent all the constraints of commercial DBMS, i.e.
primary keys, foreign key constraints, check constraints and {\em
NOT NULL}-constraints.

Our semantics is a natural extension of the semantics used in
commercial DBMSs. Note that: (a) In a DBMS there will never be a
join between a null and another value (null or not). (b) Any check
constraint with comparison, e.g $<,>, =$, will never create an
inconsistency when comparing a null value with any other value.
These two features justify our decision in Definition
\ref{def:pseudosat} to include the attributes in the joins and the
elements in $\varphi$ among the attributes that are checked to be
null with {\it IsNull}, because if there is a null in them an
inconsistency will never arise.

Our semantics of IC satisfaction with null values allows us to
integrate our results in a compatible way with current commercial
implementations; in the sense that the database repairs we will
introduce later on would be accepted as consistent by current
commercial implementations (for the classes of constraints that
can be defined and maintained by them).

\section{Repairs of Incomplete Databases}
\label{sec:repSem}

Given a database instance $D$, possibly with null values, that is
inconsistent, i.e. $D$ does not satisfy a given set $\IC$ of ICs
of the kind defined in Section \ref{sec:ICSat} or NNCs.  A {\em
repair} of $D$ will be a new instance with the same schema as $D$
that satisfies $\IC$ and minimally differs from $D$.

More formally, for database instances $D, D'$ over the same
schema, the {\it distance} between  them was defined in
\cite{ABC99} by means of the symmetric difference $\Delta(D,D')=
(D \smallsetminus D') \cup ( D' \smallsetminus D)$.
Correspondingly, a repair of $D$ wrt $\IC$ was defined as an
instance $D'$ that satisfies $\IC$ and minimizes $\Delta(D,D')$
under set inclusion. Finally, a tuple $\bar{t}$ was defined as a
consistent answer to a query $Q(\bar{x})$ in $D$ wrt $\IC$ if
$\bar{t}$ is an answer to $Q(\bar{x})$ from every repair of $D$
wrt $\IC$. The definition of repair given in \cite{ABC99}
implicitly ignored the possible presence of null values.
Similarly, in \cite{tplp,padl,cali}, that followed the repair
semantics in \cite{ABC99}, no null values were used in repairs.

\begin{example}\label{ex:repInf} Consider the database $\D$ below and the RIC: $\nit{Course(ID, Code})
\rightarrow$ \vspace*{-7mm}\begin{multicols}{2}
\noindent \begin{tabular}{c|c|c|} \hline Course & ID & Code\\
\hline
 & {\em 21} & {\em C15} \\
 & {\em 34} & {\em {C18}}\\
 \cline{2-3}
 \end{tabular}
~~~\begin{tabular}{c|c|c|} \hline Student & ID & Name\\
\hline
 & {\em 21} & {\em Ann}\\
 & {\em 45} & {\em Paul} \\
 \cline{2-3}
 \end{tabular}

\noindent $\exists \nit{Name} ~\nit{Student}(\nit{ID,Name})$. $\D$
is inconsistent, because there is no tuple in $\nit{Student}$ for
tuple {\em Course(34,C18)}  in \end{multicols} \vspace*{-4mm}
\noindent $Course$. The database can be minimally repaired by
deleting the inconsistent tuple or by inserting a new tuple into
table {\em Student}. In the latter case, since the value for
attribute {\em Name} is unknown, we should consider repairs with
all the possible values in the domain. Therefore, for  the repair
semantics introduced in \cite{ABC99}, the repairs are of the two
following forms

\noindent ~\begin{tabular}{c|c|c|} \hline ~Course~ & ID & Code\\
\hline
 & {\em 21} & {\em C15} \\
 & & \\
  & & \\
 \cline{2-3}
 \end{tabular}
~\begin{tabular}{c|c|c|} \hline ~Student~ & ID & Name\\
\hline
 & {\em 21} & {\em Ann}\\
 & {\em 45} & {\em Paul} \\
  & & \\
 \cline{2-3}
 \end{tabular}~~~
\begin{tabular}{c|c|c|}
\hline ~Course~ & ID & Code\\ \hline
 & {\em 21} & {\em C15} \\
 & {\em 34} & {\em {C18}}\\
 & & \\
 \cline{2-3}
 \end{tabular}
~\begin{tabular}{c|c|c|} \hline ~Student~ & ID & Name\\
\hline
 & {\em 21} & {\em Ann}\\
 & {\em 45} & {\em Paul} \\
 & {\em 34} & {\em  $\mathbf{\mu}$}  \\
 \cline{2-3}
 \end{tabular}

\noindent for all the possible values of $\mu$ in the domain,
obtaining a possibly infinite number of repairs. \boxtheorem
\end{example}
The problem of deciding if a tuple is a consistent answer to a
query wrt to a set of universal and referential ICs is undecidable
for this repair semantics \cite{cali}.

An alternative approach is to consider that, in a way, the value
$\mathbf{\mu}$ in Example \ref{ex:repInf} is an unknown value, and
therefore, instead of making it take all the values in the domain,
we could use it as a null value. We will pursue this idea, which
requires to modify the notion of repair accordingly. It will turn
out that consistent query answering will become decidable for
universal and referential constraints.

\begin{example} \label{ex:repInf2} (example \ref{ex:repInf}
cont.) By using null values, there will be only two repairs:
\vspace{-3mm}\begin{multicols}{2} \noindent Repair 1:

\noindent Repair 2:
\end{multicols}
\vspace{-3mm}\noindent\begin{tabular}{c|c|c|} \hline ~Course~ & ID & Code\\
\hline
 & {\em 21} & {\em C15} \\
 & & \\
 & & \\
 \cline{2-3}
 \end{tabular}
~\begin{tabular}{c|c|c|} \hline ~Student~ & ID & Name\\
\hline
 & {\em 21} & {\em Ann}\\
 & {\em 45} & {\em Paul} \\
  & & \\
 \cline{2-3}
 \end{tabular}~~~~\begin{tabular}{c|c|c|} \hline ~Course~ & ID & Code\\
\hline
 & {\em 21} & {\em C15} \\
 & {\em 34} & {\em {C18}}\\
 & & \\
 \cline{2-3}
 \end{tabular}
~\begin{tabular}{c|c|c|} \hline ~Student~ & ID & Name\\
\hline
 & {\em 21} & {\em Ann}\\
 & {\em 45} & {\em Paul} \\
 & {\em 34} & {\em  $\nn$}  \\
 \cline{2-3}
 \end{tabular}

\noindent Here $\nn$ tells us that there is a tuple with 34 in the
first attribute, but unknown value in the second. \boxtheorem
\end{example}
Now we define in precise terms the notion of repair of a database
with null values.

\begin{definition} \label{def:minD} \em \cite{book03}
Let $D,D',D''$ be database instances over the same schema and
domain $\cal U$. It holds $D' \leq_{D} D''$ iff: ~(a)~ For every
database atom $P\oa \in \Delta(D,D')$, with $\bar{a}
    \in ({\cal U} \smallsetminus \{\nn\})$,\footnote{That $\bar{a}
    \in ({\cal U} \smallsetminus \{\nn\})$ means that each of the elements in tuple $\bar{a}$ belongs
    to $({\cal U} \smallsetminus \{\nn\})$.} it holds $P\oa \in
    \Delta(D,D'')$;  and ~(b)~ For every atom
    $Q(\bar{a},\overline{\nn})\footnote{$\overline{\nn}$ is a tuple of null values, that, to simplify the presentation,
    are placed in the last attributes of $Q$, but could be anywhere else in $Q$. } \in \Delta(D,D')$, with $\bar{a}
    \in ({\cal U} \smallsetminus \{\nn\})$, there exists a $\bar{b}
    \in {\cal U}$ such that $Q(\bar{a},\bar{b}) \in \Delta(D,D'')$ and  $Q(\bar{a},\bar{b}) \not \in \Delta(D,D')$.  \boxtheorem
\end{definition}

\vspace{-3mm}
\begin{definition} \label{def:repair} \em
Given a database instance $D$ and a set $\IC$ of ICs of the form
(\ref{eq:formatGen}) and NNCs, a repair of $D$ wrt $\IC$ is a
database instance $D'$ over the same schema, such that $D'
\models_{_N} \IC$  and $D'$ is $\leq_{D}$-{\it minimal} in the
class of database instances that satisfy $\IC$ wrt $\models_{_N}$,
and share the schema with $D$, i.e. there is no database $D''$ in
this class with $D'' <_D D'$, where $D'' <_{D} D'$ means $D''
\leq_{D} D'$ but not $D' \leq_D D''$. The set of repairs of $D$
wrt $\IC$ is denoted with $\nit{Rep}(D,\IC)$. \boxtheorem
\end{definition}

\vspace{-1mm}\noindent
 In the absence of $\nn$, this definition of repair coincides with
the one  in \cite{ABC99}.

\begin{example}\label{ex:nonGeneric} The database instance
$D=\{Q(a,b),P(a,c)\}$ is inconsistent wrt the ICs  $\psi_1\!:
(P(x,y)$ $ \rightarrow$ $ \exists z Q(x,z))$ and $\psi_2\!:
(Q(x,y)$ $ \rightarrow$ $ y \not = b)$.\footnote{The second IC is
{\em non-generic} \cite{bookJan}
 in the sense that it implies some ground database literals. Non generic ICs have in general
been  left aside in the literature on CQA.} because $D \not \schn
\psi_2$. The database has two repairs wrt $\{\psi_1, \psi_2\}$,
namely $D_1=\{\}$, with $\Delta(D,D_1)=\{Q(a,b),P(a,c)\}$, and
$D_2=\{P(a,b), Q(a,\nn))\}$, with $\Delta(D,D_2)=\{Q(a,b),$
$Q(a,\nn)\}$. Notice that $D_2 \not \leq_D D_1$ because
$Q(a,\nn)\in \Delta(D,D_2)$ and there is no constant $d \in {\cal
U}$ such that $Q(a,d)\in \Delta(D,D_1)$ and
 $Q(a,d) \not \in \Delta(D,D_2)$. Similarly,  $D_1
\not \leq_D D_2$, because $P(a,c)\in \Delta(D,D_1)$ and
$P(a,c)\not \in \Delta(D,D_1)$.  \boxtheorem
\end{example}

\begin{example}
If the database instance is $\{P(a,\nn),P(b,c),R(a,b)\}$ and $\IC$
consists only of $(P(x,y) \rightarrow \exists z~ R({x},$ $z))$,
then there are two repairs: $D_1=\{P(a,\nn),$ $P(b,c),$ $ R(a,b),$
$ R(b, \nn)\}$, with $\Delta(D,D_1)=\{R(b,\nn)\}$, and
$D_2=\{P(a,\nn),$ $ R(a,b)\}$, with $\Delta(D,D_2)=\{P(b,c)\}$.
Notice, for example, that $D_3=\{P(a,\nn),$ $P(b,c),$ $ R(a,b),$ $
R(b,d)\}$, for~ any $d \in {\cal U}$ different from $\nn$, is not
a repair: Since $\Delta(D,D_3)=\{R(b,d)\}$, we have $D_2 <_D D_3$
and, therefore  $D_3$ is not $\leq_D$-minimal.
 \boxtheorem
\end{example}

\begin{example}\label{ex:repCyc} Consider the UIC
$\forall xy(P(x,y) \rightarrow T(x))$ and  the RIC $\forall
x(T(x)\rightarrow \exists y P(y,x))$, and the inconsistent
database $D=\{P(a,b),P(\nn,a),T(c)\}$. In this case, we have a
RIC-cyclic set of ICs. The four repairs are

\vspace*{-3mm}\begin{center}
\begin{tabular}{c|c|c}
    $i$& $D_i$ & $\Delta(D,D_i)$\\
    \hline
        1 & $\{P(a,b),P(\nn,a),T(c),P(\nn,c),T(a)\}$ & $\{T(a),P(\nn,c)\}$\\
        2 & $\{P(a,b),P(\nn,a),T(a)\}$ & $\{T(a),T(c)\}$\\
        3 & $\{P(\nn,a),T(c),P(\nn,c)\}$ &  $\{P(a,b),P(\nn,c)\}$\\
        4 & $\{P(\nn,a)\}$ & $\{P(a,b),T(c)\}$
\end{tabular}
\end{center}

\vspace*{-2mm}\noindent Notice that, for example, the additional
instance $D_5=\{P(a,b), $ $P(\nit{\nn},a),$ $T(c),$ $ P(c,a),
T(c)\}$, with $\Delta(D,D_5)=\{T(a),P(c,a)\}$, satisfies $\IC$,
but is not a repair because $D_1 <_{D} D_5$. \boxtheorem
\end{example}
The previous example shows that we obtain a finite number of
repairs (with finite extension). If we repaired the database by
using the non-null constants in the infinite domain with the
repair semantics of \cite{ABC99}, we would obtain an infinite
number of repairs and infinitely many of them with infinite
extension, as considered in \cite{cali}.

\begin{example} \label{ex:repair} Consider a schema with relations $R(X,Y)$, with primary
key $R[1]$, and a table $S(U,V)$, with $S[2]$ a foreign key  to
table $R$. The ICs are $\forall xyz ~(R(x,y)$ $\wedge R(x,z)
\rightarrow y=z)$ and $\forall uv~(S(u,v) \rightarrow \exists
y~R(v,y))$, plus the NNC $\forall xy (R(x,y) \wedge
\nit{IsNull}(x) \rightarrow {\bf false})$. Since the original
database satisfies the NNC and there is no constraint with an
existential quantifier over $R[1]$, the NNC will not be violated
while trying to solve other inconsistencies. We would have a {\em
non-conflicting interaction} of RICs and NNCs. Here $D =
\{R(a,b),R(a,c),S(e,f),$ $S(\nn,a)\}$ is inconsistent and its
repairs are $D_1=\{R(a,b),S(e,f),S(\nn,a),R(f,\nn)\}$,
$D_2=\{R(a,c),S(e,f),S(\nn,a),R(f,\nn)\}$,
$D_3=\{R(a,b),S(\nn,a)\}$ and $D_4=\{R(a,c),S(\nn,a)\}$\boxtheorem
\end{example}
If a given database $D$ is consistent wrt a set of ICs, then there
is only one repair, that coincides with $D$. The following example
shows what can happen if we have a {\em conflicting interaction}
of a RIC containing an existential quantifier over a variable with
an additional NNC that prevents that variable from taking null
values.

\begin{example} \label{ex:NNC} Consider  the database $\D=\{P(a),P(b),Q(b,c)\}$,
the RIC $\forall x~(P(x)$ $\rightarrow$ $\exists y ~Q(x,y))$, and
the NNC $\forall xy(Q(x,y) \wedge \nit{IsNull}(y) \rightarrow {\bf
false})$ over  an existentially quantified attribute in the RIC.
We cannot repair as expected using null values. Actually, the
repairs are $\{P(b),Q(b,c)\}$, corresponding to a tuple deletion,
but also those of the form $\{P(a),P(b),Q(b,c),Q(a,$ $\mu)\}$, for
every $\mu \in ({\cal U} \smallsetminus \{\nn\})$, that are
obtained by  tuple insertions. We thus recover the repair
semantics of \cite{ABC99}. \boxtheorem
\end{example}

\noindent With an appropriate conflicting interaction of RICs and
NNCs we could recover in our setting the situation where
infinitely many repairs and infinitely many with finite extension
appear (c.f. remark after Example \ref{ex:repCyc}). Our repair
semantics above could be modified in order to repair only through
tuple deletions in this case,  when null values cannot be used due
to the presence of conflicting NNCs. This could be done as
follows: If $\nit{Rep}(D,\IC)$ is the class of repairs according
to Definitions \ref{def:minD} and \ref{def:repair}, the
alternative class of repairs, $\nit{Rep}_d(D,\IC)$, that prefers
tuple deletions over insertions with arbitrary  non-null elements
of the domain due to the presence of conflicting NNCs, can be
defined by $\nit{Rep}_d(D,\IC):=$ $ \{\D' ~|~ D' \in
\nit{Rep}(D,\IC)$ and there is no $D'' \in \nit{Rep}(D,\IC')
\mbox{ with } D'' <_D D'\},$ where $\IC'$ is $\IC$ without the
(conflicting) NNCs.

 Since
the semantics introduced Definitions \ref{def:minD} and
\ref{def:repair} is easier to deal with, and in order to avoid
repairs like those in  Example \ref{ex:NNC}, we will make the
following

\vspace{2mm} \noindent {\bf Assumption:}~ Our sets $\IC$,
consisting of ICs of the form (\ref{eq:formatGen}) and NNCs, are
{\em non-conflicting}, in the sense that there is no NNC on an
attribute that is existentially quantified in an IC of the form
(\ref{eq:formatGen}).

\vspace{2mm} In this way, we will always be able to repair RICs by
tuple deletions or tuple insertions with null values. Notice that
every set of ICs consisting of primary key constraints (with the
keys set to be non-null), foreign key constraints, and check
constraints satisfies this condition. Also note that if there are
non conflicting NNCs, the original semantics and the one based on
$\nit{Rep}_d$-repairs coincide. The repair programs introduced in
Section \ref{sec:repProg} compute specify the
$\nit{Rep}_d$-repairs, so our assumption is also relevant from the
computational point of view.

Notice that with our repair semantics, we can prove that there
will always exists a repair for a database $D$ and a set of
non-conflicting constraints ICs; and that the set of repairs is
finite and each of them is finite in extension (i.e. each database
relation is finite), because a database instance with no tuples
always satisfies the constraints, and the domain of the repairs
can be restricted to $\nit{adom}(D) \cup \nit{const}(\IC) \cup
\{\nn\}$, where $\nit{adom}(D)$ is the active domain of
 the original instance $D$ and $\nit{const}(\IC)$ is the set of constants that
appear in the constraints.

\begin{proposition}\em \label{prop:finite} Given a
database $\D$ and a set $\IC$ of non-conflicting ICs: (a) For
every repair $D' \in {\nit Rep}(\D,$ $\IC)$, $\nit{adom}(D')
~\subseteq~ \nit{adom}(D) \cup \nit{const}(\IC) \cup \{\nn\}$.\\
(b) The set ${\nit Rep}(\D,\IC)$ of repairs is non-empty and
finite; and every $\D' \in {\nit Rep}(D,\IC) $ is
finite.\footnote{For proofs of all results
 go to {\tt www.scs.carleton.ca/$^\sim$lbravo/IIDBdemos.pdf}} \boxtheorem
\end{proposition}

\vspace*{-2mm}\begin{theorem}\label{theo:coNP} \em The problem of
determining if a database $D'$ is a repair of $D$ wrt a set $\IC$
consisting of ICs of the form (\ref{eq:formatGen}) and
NNCs\footnote{In this case we do not need the assumption of
non-conflicting ICs}  is $\nit{coNP}$-complete. \boxtheorem
\end{theorem}

\vspace*{-2mm}\begin{definition} \em \cite{ABC99}
\label{def:consistentanswer} Given a
 database $\D$, a set of  ICs $\IC$, and a  query $Q(\bar{x})$,  a
  ground tuple $\bar{t}$ is a {\it consistent answer} to
 $Q$ wrt $\IC$ in $D$ iff  for every
  $\D' \in {\nit Rep}(\D,\IC)$, ~$D' \models  Q[\bar{t}]$. If $Q$ is a sentence
  (boolean query), then $\nit{yes}$ is a consistent answer iff $\D' \models
  Q$ for every $\D' \in \nit{Rep}(D,\IC)$. Otherwise, the consistent answer is $\nit{no}$.
  \boxtheorem
 \end{definition}
In this formulation of CQA we are using a notion $D' \models
Q[\bar{t}]$ of satisfaction of queries in a database with null
values. At this stage, we are not committing to  any particular
semantics for query answering in this kind of databases. In the
rest of the paper, we will assume that we have such a notion, say
$\models_{N}^q$, that can be applied to queries in databases with
null values. Some proposals can be found in the literature
\cite{SQL2003,LL99b,Zan84}. In principle, $\models_N^q$ may be
orthogonal to the notion $\models_N$ for satisfaction of ICs.
However, in the extended version of this paper we will present a
semantics for query answering that is compatible with the one for
IC satisfaction. For the moment we are going to assume that
$\models_N^q$ can be computed in polynomial time in data for safe
first-order queries, and that it coincides with the classical
first-order semantics for queries and databases without null
values. We will also assume in the following that queries are safe
\cite{GT87}, a sufficient syntactic condition for domain
independence.

\noindent The decision problem of consistent query answering is
\vspace{-2mm}$$\nit{CQA}(Q,\IC) = \{(D,\bar{t})~|~ \bar{t} \mbox{
is a
 consistent answer to } Q(\bar{x}) \mbox{ wrt } \IC \mbox{ in }
 D\}.$$

 \vspace{-2.5mm} \noindent Since we have $Q$ and $\IC$ as parameters of the problem,
 we are interested in the data complexity of this problem, i.e. in
 terms of the size of the database \cite{AHV95}.
It turns out that  CQA for  FOL queries  is decidable, in contrast
to what happens with the classic repair semantics \cite{ABC99}, as
established in \cite{cali}.

\begin{theorem} \label{theo:decidable} \em Consistent
query answering for  first-order queries wrt to non-conflicting
sets of ICs of the form (\ref{eq:formatGen})  and NNCs  is
decidable. \boxtheorem
\end{theorem}
\vspace{-1mm}The ideas behind the proof are as follows: (a) There
is a finite number of database instances that are candidates to be
repair given that the use only the active domain of the original
instance, $\nn$ and the constants in the ICs. (b) The satisfaction
of ICs in the candidates can de decided by restriction to the
active domain given that the ICs are domain independent.  (c)
Checking if $D_1 \leq_D D_2$ can be effectively decided. (d) The
answers to safe first-order queries can be effectively computed.

The following proposition can be obtained by using a similar
result \cite{cm-2005} and the fact that our tuple deletion based
repairs are exactly those considered in \cite{cm-2005}, and every
repair in our sense that is not one of those contains at least one
tuple insertion.

\vspace{-2mm}\begin{theorem}\label{theo:Pi2p} \em Consistent query
answering for first-order queries and non-conflict-ing sets of ICs
of the form (\ref{eq:formatGen}) or NNCs is $\Pi_2^p$-complete.
\boxtheorem
\end{theorem}
In the proof of this theorem  NNCs are not needed for hardness.
Actually, hardness can be obtained with boolean queries.

\vspace{-3mm}
\section{Repair Logic Programs} \label{sec:repProg}

\vspace{-1mm}The stable models semantics was introduced in
\cite{glELPb} to give a semantics to disjunctive logic programs
that are non-stratified, i.e. that contain recursive definitions
that contain weak negation. By now it is the standard semantics
for such programs. Under this semantics, a program may have
several stable models; and what is true of the program is what is
true in all its stable models (a cautious semantics).

Repairs of relational databases can be specified as stable models
of disjunctive logic programs. In \cite{book03,cascon,sccc05} such
programs were presented, but they were based on classic IC
satisfaction, that differs from the one introduced in Section
\ref{sec:ICSat}.

The repair programs we will present now implement the repair
semantics introduced in Section \ref{sec:ICSat} for a set of
RIC-acyclic constraints. The repair programs use annotation
constants with the intended, informal semantics shown in the table
below. The annotations are used in an extra attribute introduced
in each database predicate; so for a predicate $P \in {\cal R}$,
the new version of it, $P\!\!\_$~, contains an extra attribute.

\vspace{-3mm}\begin{center}
\begin{tabular}{|l|l|l|}
\hline
~Annotation &  ~Atom & ~The tuple $P(\bar{a})$ is...\\
\hline
~~$\ta$ & ~$P\!\!\_(\bar{a},\ta)$& ~advised to be made true\\
~~$\fa$ & ~$P\!\!\_(\bar{a},\fa)$& ~advised to be made false\\
~~$\tr^\star$ & ~$P\!\!\_(\bar{a},\tr^\star)$&  ~true or becomes true\\
~~$\tr^{\star\star}$ & ~$P\!\!\_(\bar{a},\tr^{\star\star})$&  ~it is true in the repair\\
\hline
\end{tabular}
\end{center}
\vspace{-3mm}In the repair program,  $\nn$ is treated as any other
constant in ${\cal U}$, and therefore the $\nit{IsNull}(x)$ atom
can be replaced by $x = \nn$.

\begin{definition}\label{def:optrp} \em Given a
database instance $\D$, a set $\IC$ of UICs, RICs and NNCs, the
repair program $\Pi(\D,$ $\IC)$ contains the following rules:
\vspace*{-2mm}\begin{enumerate} \item \label{it:fact}Facts:
$P(\bar{a})$ for each atom $P(\bar{a}) \in \D$.
\item \label{it:uic} For every UIC $\psi$ of  form (\ref{eq:format}), the rules:\\
{\small $\bigvee_{i=1}^{n} P\!\!\_ _{i}(\bar{x}_{i},\fa) \vee
\bigvee_{j=1}^{m} Q\!\!\_ _{j}(\bar{y}_{j}, \mathbf{t_{a}})
~\leftarrow~ \bigwedge_{i=1}^{n} P\!\!\_ _{i}(\bar{x}_{i},\trs),
\bigwedge_{Q\!\!\_ _{j} \in Q'} Q\!\!\_ _{j}(\bar{y}_{j},\fa),$\\
\hspace*{5.4cm} $\bigwedge_{Q_{k} \in Q''}$ $\n
Q_{k}(\bar{y}_{k}), \bigwedge_{x_l \in {\cal A}(\psi) \cap
\bar{x}}$ $x_l\neq \nn , \bar{\varphi}.$}\\for every set $Q'$ and
$Q''$ of atoms appearing in formula (\ref{eq:format}) such that
$Q' \cup Q'' = \bigcup_{j=1}^m Q_{j}(\bar{y_j})$ and $Q' \cap Q''
= \emptyset$.\footnote{We are assuming in this definition that the
rules are a direct translation of the original ICs introduced in
Section \ref{sec:prelim}; in particular, the same variables are
used and the standardization conditions  about their occurrences
are respected in the program.} Here ${\cal A}(\psi)$ is the set of
relevant attributes for $\psi$, $\bar{x}=\bigcup_{i=1}^n x_i$ and
$\bar{\varphi}$ is a conjunction of built-ins that is equivalent
to the negation of $\varphi$.

\item \label{it:ric} For every RIC  of  form (\ref{eq:formatRIC}),
the rules:\\ {\small $P\!\!\_(\bar{x},\mathbf{f_a}) \vee
Q\!\!\_(\bar{x}',\overline{\nn},\mathbf{t_a}) \leftarrow
P\!\!\_(\bar{x},\mathbf{t^{\star}}),  \n ~\nit{aux}(\bar{x}'),
 \bar{x}'\neq \nn.$\\} And for every {\small $y_i \in \bar{y}$}:\\
{\small $\nit{aux}(\bar{x}') \leftarrow
Q\!\!\_(\bar{x}',\bar{y},\trs),$ $\n
Q\!\!\_(\bar{x}',\bar{y},\fa),$ $\bar{x}'\neq \nn,$ $y_i \neq
\nn$.}

\item \label{it:nnc} For every NNC of  the form (\ref{eq:SatruleNNC}), the rule:\\
{\small $P\!\!\_(\bar{x},\mathbf{f_a}) \leftarrow
P\!\!\_(\bar{x},\mathbf{t^{\star}}),  x_i=\nn.$}
\item \label{it:ts}For each predicate $P \in R$, the  annotation rules:\\
{\small $P\!\!\_(\bar{x},\trs) \leftarrow P(\bar{x}).$
\hspace{1cm} $P\!\!\_(\bar{x},\trs) \leftarrow
P\!\!\_(\bar{x},\ta).$}

\item \label{it:tss}For every predicate $P \in {\cal R}$, the
interpretation rule:\\
{\small $P\!\!\_(\bar{x},\tss)$ $\leftarrow
P\!\!\_(\bar{x},\trs),$ $~not~P\!\!\_(\bar{x},\fa).$}

\item \label{it:dc}For every predicate $P \in {\cal R}$,  the
program denial constraint:~\\
{\small $\leftarrow~P\!\!\_(\bar{x},\ta),$
$P\!\!\_(\bar{x},\fa).$} \boxtheorem
\end{enumerate}
\end{definition}
\vspace*{-2mm}Facts in 1. are the elements of the database. Rules
2., 3. and 4. capture, in the right-hand side, the violation of
ICs of the forms (\ref{eq:format}), (\ref{eq:formatRIC}), and
(\ref{eq:SatruleNNC}), resp., and, with the left-hand side, the
intended way of restoring consistency. The set of predicates $Q'$
and $Q''$ are used to check that in all the possible combinations,
the consequent of a UIC is not being satisfied.
 Since the
satisfaction of UICs and RICs needs to be checked only if none of
the relevant attributes of the antecedent are $\nn$, we use $x
\not = \nn$ in rule 2. and in the first two rules in 3. (as usual,
$\bar{x}' \neq \nn$ means the conjunction of the atoms $x_j \neq
\nn$ for $x_j \in \bar{x}'$). Notice that rules 3. are implicitly
based on the fact that the relevant attributes for a RIC of the
form (\ref{eq:formatRIC}) are ${\cal A} = \{x ~|~ x \in
\bar{x}'\}$. ~Rules 5. capture the atoms that are part of the
inconsistent database or that become true in the repair process;
and rules 6. those that become true in the repairs. Rule 7.
enforces, by discarding models, that no atom can be made both true
and false in a repair.

\begin{example} \label{ex:repair2} (example \ref{ex:repair} cont.)
The repair program $\Pi(\dbic)$ is the following:

{\small
\noindent 1.~ $R(a,b).$~~~~$R(a,c).$~~~~$S(e,f).$~~~~$S(\nn,a).$\\
2.~
 $R\!\!\_(x,y,\mathbf{f_a}) \vee R\!\!\_(x,z,\mathbf{f_a}) ~\leftarrow~
R\!\!\_(x,y,\mathbf{t^{\star}}),R\!\!\_(x,z,\mathbf{t^{\star}}),$ $y \not = z,$ $x \not = \nn.$\\
3.~ $S\!\!\_(u,x,\mathbf{f_a}) \vee
R\!\!\_(x,\nit{\nn},\mathbf{t_a}) \leftarrow
S\!\!\_(u,x,\mathbf{t^{\star}}), \mathit{not}~aux(x),$ $x \not = \nn.$\\
\hspace*{5mm}$aux(x) \leftarrow  R\!\!\_(x,y,\trs),$ $\n R\!\!\_(x,y,\fa),$ $x \not = \nn, $ $y \not = \nn.$\\
5.~ $R\!\!\_(x,y,\mathbf{t^{\star}}) \leftarrow
R\!\!\_(x,y,\mathbf{t_a}).
\hspace*{3mm}R\!\!\_(x,y,\mathbf{t^{\star}}) \leftarrow
R\!\!\_(x,y,\mathbf{t_d}).$~ {\hfill (similarly for $S$)}\\
 6. ~$R\!\!\_(x,y,\mathbf{t^{\star\star}})
\leftarrow R\!\!\_(x,y,\mathbf{t_a}).$\\
\hspace*{5mm}$R\!\!\_(x,y,\mathbf{t^{\star\star}}) \leftarrow
R(x,y),~\mathit{not}~R\!\!\_(x,y,\mathbf{f_a}).$
 \hspace*{5mm}{\hfill (similarly for $S$)}\\
7.~ $\leftarrow R\!\!\_(x,y,\mathbf{t_a}),
R\!\!\_(x,y,\mathbf{f_a}). ~~~ \leftarrow
S\!\!\_(\bar{x},\mathbf{t_a}), S\!\!\_(\bar{x},\mathbf{f_a}).$}

\vspace{2mm}\noindent Only rules 2. and 3. depend on the ICs:
rules 2. for the UIC, and 3. for the RIC. They say how to repair
the inconsistencies. In rule 2., $Q'=Q''=\emptyset$, because there
is no database predicate in the consequent of the UIC. There is no
rule 4., because there is no NNC. \boxtheorem
\end{example}

\begin{example}
Consider $D=\{P(a,b),P(c,\nn)\}$ and the non-conflicting set of
ICs: $\{\forall P(x,y) \rightarrow R(x) \vee S(y)$,$P(x,y) \wedge
\nit{IsNull}(y) \rightarrow \mathbf{false}\}$. Then
$\Pi(\D,\IC):$\\
{\small \noindent 1.~ $P(a,b). ~~~~ P(c,\nn).$\\
2.~ $P\!\!\_(x,y,\fa)  \vee  R\!\!\_(x,\ta) \vee S\!\!\_(y,\ta)
\leftarrow P\!\!\_(x,y,\trs), R\!\!\_(x,\fa), $ $S\!\!\_(y,\fa),$
$x\neq \nn, y\neq \nn.$\\
\hspace*{4.5mm}$P\!\!\_(x,y,\fa)  \vee  R\!\!\_(x,\ta) \vee
S\!\!\_(y,\ta) \leftarrow P\!\!\_(x,y,\trs), R\!\!\_(x,\fa), $ $\n
S(y),$ $x\neq \nn, y\neq \nn.$\\
\hspace*{4.5mm}$P\!\!\_(x,y,\fa)  \vee  R\!\!\_(x,\ta) \vee
S\!\!\_(y,\ta) \leftarrow P\!\!\_(x,y,\trs), \n
R(y),$ $ S\!\!\_(x,\fa), $  $x\neq \nn, y\neq \nn.$\\
\hspace*{4.5mm}$P\!\!\_(x,y,\fa)  \vee  R\!\!\_(x,\ta) \vee
S\!\!\_(y,\ta) \leftarrow P\!\!\_(x,y,\trs),$ $ \n R(y),$ $ \n
S(y),$ $x\neq \nn, y\neq \nn.$\\
4.~ $P\!\!\_(x,y,\fa) \leftarrow P\!\!\_(x,y,\trs), y = \nn.$\\
5.~ $P\!\!\_(x,y,\trs) \leftarrow P\!\!\_(x,y,\ta).$
\hspace*{5mm}$ P\!\!\_(x,y,\trs) \leftarrow P(x,y).$  {\hfill (similarly for $R$ and $S$)}\\
6.~ $P\!\!\_(x,y,\tss) \leftarrow P\!\!\_(x,y,\ta).$\\
\hspace*{4.5mm}$ P\!\!\_(x,y,\tss) \leftarrow P(x,y), \n P\!\!\_(x,y,\fa).$ {\hfill (similarly for $R$ and $S$)}\\
7.~ $\leftarrow P\!\!\_(x,y,\ta), P\!\!\_(x,y,\fa).$ {\hfill
(similarly for $R$ and $S$)}}\\ The rules in 2. are constructed by
choosing all the possible sets $Q'$ and $Q''$ such that $Q'\cup
Q''=\{R(x),S(y)\}$ and $Q' \cap Q''=\emptyset$. The first rule in
2. corresponds to $Q'=\{R(x),S(y)\}$ and $Q''=\emptyset$, the
second for $Q'=\{R(x)\}$ and $Q''=\{S(y)\}$, the third for
$Q'=\{S(y)\}$ and $Q''=\{R(x)\}$, and the fourth for
$Q'=\emptyset$ and $Q''=\{R(x),S(y)\}$ \boxtheorem
\end{example}

 \vspace{-1mm}\noindent The repair program can be run by a logic
programming system that computes the stable models semantics, e.g.
DLV system \cite{leoneCorr}.  The repairs can be obtained by
collecting the atoms annotated with $\tss$ in the stable models of
the program.

\vspace{-1mm}\begin{definition} \label{def:inst} \em Let $\cal M$
be a stable model of program  $\Pi(\dbic)$.  The database instance
associated with $\cal M$ is ~$D_{\cal M}~ =~ \{P\oa \mid P \in
{\cal R} \mbox{ and } P\oatstarr \in {\cal M}\}$. \boxtheorem
\end{definition}

\vspace*{-2mm}\begin{example}\label{ex:repair3} (example
\ref{ex:repair2} continued) The program has four stable models
(the facts of the program are omitted for simplicity):

{\small \noindent \begin{tabular}{cp{11.1cm}} $\mm_1=$ &
$\{R\_(a,b,\trs),$ $ R\_(a,c,\trs),$ $ S\_(e,f,\trs),$ $
S\_(\nn,a,\trs),$ $ aux(a),$ $ \underline{S\_(e,f,\tss)},$ $
\underline{S\_(\nn,a,\tss)},$ $ R\_(f,\nn,\ta),$ $
\underline{R\_(a,b,\tss)},$ $
R\_(a,c,\fa),$ $ R\_(f,\nn,\trs),$ $ \underline{R\_(f,\nn,\tss)}$ $\}$,\\
\end{tabular}

\noindent \begin{tabular}{cp{11.1cm}} $\mm_2=$ &$\{R\_(a,b,\trs),$
$ R\_(a,c,\trs),$ $ S\_(e,f,\trs),$ $ S\_(\nn,a,\trs),$ $ aux(a),$
$ \underline{S\_(e,f,\tss)},$ $ \underline{S\_(\nn,a,\tss)},$ $
R\_(f,\nn,\ta),$ $ R\_(a,b,\fa),$ $ \underline{R\_(a,c,\tss)},$ $
R\_(f,\nn,\trs),$ $ \underline{R\_(f,\nn,\tss)}$ $\}$,
\end{tabular}

\noindent \begin{tabular}{cp{11.1cm}}$\mm_3=$ & $\{R\_(a,b,\trs),$
$ R\_(a,c,\trs),$ $ S\_(e,f,\trs),$ $ S\_(\nn,a,\trs),$ $ aux(a),$
$ S\_(e,f,\fa),$ $ \underline{S\_(\nn,a,\tss)},$ $
\underline{R\_(a,b,\tss)},$ $ R\_(a,c,\fa)\}$,
\end{tabular}

\noindent \begin{tabular}{cp{11.1cm}} $\mm_4=$ &
$\{R\_(a,b,\trs),$ $ R\_(a,c,\trs),$ $ S\_(e,f,\trs),$ $
S\_(\nn,a,\trs),$ $ aux(a),$ $ S\_(e,f,\fa),$ $
\underline{S\_(\nn,a,\tss)},$ $ R\_(a,b,\fa),$ $
\underline{R\_(a,c,\tss)}\}$.
\end{tabular}
}

\noindent The databases associated to the models select the
underlined atoms: $D_1= \{S(e,f),$ $S(\nn,a),$ $R(a,b),$ $
R(f,\nn)\}$, $D_2= \{S(e,f),$ $S(\nn,a),$ $R(a,c),$ $R(f,$
$\nn)\}$ $D_3= \{S(\nn,a),R(a,b)\}$ and $D_4=
\{S(\nn,a),R(a,c)\}$. As expected these are the repairs obtained
in Example \ref{ex:repair}. \boxtheorem
\end{example}

\vspace{-2mm}\begin{theorem}\em \label{theorem:corresp2} Let $\IC$
be a RIC-acylic set of UICs, RICs and NNCs. If $\mm$ is a stable
model of $\Pi(\dbic)$, then $D_{\mm}$ is a repair of $D$ with
respect to $\IC$.  Furthermore, the repairs obtained in this way
are all the repairs of $D$. \boxtheorem
\end{theorem}

\section{Head-Cycle-Free Programs}\label{sec:HCFRIC}

In some cases, the repair programs introduced
 in Section \ref{sec:repProg} can be
 transformed into equivalent non-disjunctive programs. This is the case
when they become {\em head-cycle-free} \cite{rachel}. Query
evaluation from such programs has lower computational complexity
than general disjunctive programs, actually the data complexity is
reduced from $\Pi^P_2$-complete to $\nit{coNP}$-complete
\cite{rachel,voronkov}. We briefly recall their definition.

 The {\it dependency
graph}
 of a ground disjunctive program $\Pi$ is
the directed graph that has ground atoms as vertices, and an edge
from atom $A$ to atom $B$ iff there is a rule with $A$ (positive)
in the body and $B$ (positive) in the head. $\Pi$ is {\it
head­-cycle free} (HCF) iff its dependency graph does not contain
any directed cycles passing through two atoms in the head of the
same rule. A  disjunctive program $\Pi$ is HCF if its ground
version is HCF.

A HCF program  $\Pi$ can be transformed into a non-disjunctive
normal program $\nit{sh}(\Pi)$ that has the same stable models. It
is obtained by replacing every disjunctive rule of the form ~$
\bigvee_{i=1}^n P_i(\bar{x}_i) \leftarrow \bigwedge_{j=1}^m
Q_j(\bar{y}_j),~ \varphi.$~ by the $n$ rules ~$P_i(\bar{x}_i)
\leftarrow \bigwedge_{j=1}^m Q_j(\bar{y}_j),~ \varphi,~
\bigwedge_{k \not =i} ~\mathit{not}~ P_k(\bar{x}_k).$, ~for
$i=1,...,n$.

For certain classes of queries and ICs, consistent query answering
has a data complexity lower than $\Pi^P_2$, a sharp lower bound as
seen in Theorem \ref{theo:Pi2p} (c.f. also \cite{cm-2005}). In
those cases, it is natural to consider this kind of
transformations of the disjunctive repair program. In the rest of
this section we will consider  sets $\IC$ of integrity constraints
formed by UICs, RICs and NNCs.

\vspace*{-1mm}\begin{definition} \em A predicate $P$ is {\em
bilateral} with respect to $\IC$ if it belongs to the antecedent
of a constraint $ic_1 \in \IC$ and to the consequent of a
constraint $ic_2 \in \IC$, where $ic_1$ and $ic_2$ are not
necessarily different. \boxtheorem \end{definition}

\vspace*{-3mm}\begin{example} If $\IC=\{\forall x~(T(x)
\rightarrow \exists \,y~ R(x,y), \forall xy~(S(x,y) \rightarrow
T(x))\}$, the only bilateral predicate is $T$. \boxtheorem
\end{example}

\vspace*{-3mm}\begin{theorem} \em \label{theorem:HCF2} For a set
$\IC$ of UICs, RICs and NNCs, if for every $\nit{ic} \in \IC$, it
holds that (a) $\nit{ic}$ has no bilateral predicates; or (b)
$\nit{ic}$ has exactly one occurrence of a  bilateral predicate
(without repetitions), then the program $\Pi(\D,\IC)$ is HCF.
\boxtheorem
\end{theorem}
 For example, if in $\IC$ we have the constraint $P(x,y)
\rightarrow P(y,x)$, then $P$ is a bilateral predicate, and the
condition in the theorem is not satisfied. Actually, the program
$\Pi(D,\IC)$ is not HCF. If we have instead $P(x,a) \rightarrow
P(x,b)$, even though the condition is not satisfied, the program
is HCF. Therefore, the condition is sufficient, but not necessary
for the program to be HCF.

This theorem can be immediately applied to useful classes of ICs,
like denial constraints, because they do not have any bilateral
literals, and in consequence, the repair program is HCF.

\vspace*{-2mm}\begin{corollary}\em \label{cor:ric} If $\IC$
contains only  constraints of the form $
    \bar{\forall}(\bigwedge_{i=1}^{n} \!\!P_{i}(\bar{t_{i}}) ~\rightarrow~ \varphi)$,  where
    $P_{i}(\bar{t_{i}})$ is a database atom and $\varphi$ is a formula
    containing built-in predicates only, then $\Pi(D,IC)$  is HCF. \boxtheorem
\end{corollary}
As a consequence of this corollary we obtain, for first-order
queries and this class of ICs, that CQA  belongs to ${\it coNP}$,
because a query program (that is non-disjunctive) together with
the repair program is still HCF.  For this class of constraints,
with the classical tuple-deletion based semantics, this problem
becomes ${\it coNP}$-complete \cite{cm-2005}. Actually, CQA for
this class with our tuple-deletion/null-value based semantics is
still ${\it coNP}$-complete, because the same reduction found in
\cite{cm-2005} can be used in our case.

\vspace{-3mm}\section{Conclusions}

We have introduced a new repair semantics that considers,
systematically and for the first time, the possible occurrence of
null values in a database in the form we find them present and
treated in current commercial implementations. Null values of the
same kind are also used to restore the consistency of the
database. The new semantics applies to a wide class of ICs,
including cyclic sets of referential ICs.

We established the decidability of CQA under this semantics, and a
tight lower and upper bound was presented. The repairs under this
semantics can be specified as stable models of a disjunctive logic
program with a  stable model semantics for acyclic foreign key
constraints, universal ICs and {\em NOT NULL}-constraints,
covering all the usual ICs found in database practice.

In an extended version of this paper we will provide: (a) An
extension of our semantics of IC satisfaction in databases with
null values that can also be applied to query answering in the
same kind of databases. (b) A more detailed analysis of the way
null-values are propagated in a controlled manner, in such a way
that no infinite loops are created. (c) Construction of repairs
based on a sequence of ``local" repairs for the individual ICs.

 \vspace{1mm} \noindent {\bf Acknowledgments:}~~ Research
supported by NSERC, CITO/IBM-CAS Student Internship Program. ~L.
Bertossi is Faculty Fellow of  IBM Center for Advanced Studies
(Toronto Lab.).

\end{document}